\documentclass[aps,pra,groupedaddress,amsfonts,amssymb,twocolumn,final]{revtex4}
\def\eat#1{}
\def\ifundefined#1{\expandafter\ifx\csname
                        \expandafter\eat\string#1\endcsname\relax}
\ifundefined\pacs
\newenvironment{references}{\begin{thebibliography}{123}}{\end{thebibliography}}
\fi

\def\H{{\cal H}}\def\N{{\cal N}}

\def\rhom{\overline\rho}%mean rho
\def\rhoq{\rho_q}

\def\idty{{\rm 1\mkern-5.4mu I}}
\def\Ibb #1{{\rm I\mkern -3.6mu#1}}
\def\ibb #1{\hbox{\kern.3em\vrule
     height 1.5ex depth -.1ex width .2pt\kern-.3em\rm#1}}
  \def\Rl{{\Ibb R}}

\def\Ree{{\Re\mkern-2mu e}}
\def\Imm{{\Im\mkern-2mu m}}
\def\tr{{\rm tr}}

\newcommand{\bra}[1]{\mbox{$\langle #1 |$}}
\newcommand{\ket}[1]{\mbox{$| #1 \rangle$}}

\newcommand{\kettbra}[1]{\ket{#1}\!\bra{#1}}

\begin{document}
\title{The Quantum Monty Hall Problem}

\author{G.M.~D'Ariano}
\email{dariano@pv.infn.it}
%\homepage{homepage}
\affiliation{Quantum Optics and Information Group, Instituto Nazionale de Fisica della Materia, Unita di Pavia\\
  Dipartimento di Fisica ``A Volta'' via Bassi 6, I-27100 Pavia, Italy}

\author{R.D.~Gill}
\email{gill@math.uu.nl}
\homepage{http://www.math.uu.nl/people/gill}
\altaffiliation{also affiliated to EURANDOM, Netherlands}
\affiliation{Dept.~of Mathematics, University of Utrecht, 
  Box 80010, 3508 TA Utrecht, Netherlands.}

\author{M.~Keyl}
\email{M.Keyl@tu-bs.de}
\homepage{http://www.imaph.tu-bs.de/home/keyl}
\affiliation{Inst. Math. Phys., TU-Braunschweig, Mendelssohnstra{\ss}e 3,
  D-38106 Braunschweig, Germany}

\author{B.~K{\"u}mmerer}
\email{kuem@mathematik.uni-stuttgart.de}
%\homepage{homepage}
\affiliation{Math. Inst. A, Univ. Stuttgart, Pfaffenwaldring 57,
  D-70569 Stuttgart, Germany}

\author{H.~Maassen}
\email{maassen@sci.kun.nl}
%\homepage{homepage}
\affiliation{Subfaculteit Wiskunde, University of Nijmegen,
  Toernooiveld 1, 6525 ED Nijmegen, The Netherlands}

\author{R.F.~Werner}
\email{R.Werner@tu-bs.de}
\homepage{http://www.imaph.tu-bs.de/home/werner}
\affiliation{Inst. Math. Phys., TU-Braunschweig, Mendelssohnstra{\ss}e 3,
  D-38106 Braunschweig, Germany}

\begin{abstract}
We consider a quantum version of a well-known statistical decision
problem, whose solution is, at first sight, counter-intuitive to
many. In the quantum version a continuum of possible choices
(rather than a finite set) has to be considered. It can be phrased
as a two person game between a player P and a quiz master Q. Then
P always has a strategy at least as good as in the classical case, while
Q's best strategy results in a game having the same value as the
classical game. We investigate the consequences of Q storing his
information in classical or quantum ways. It turns out that Q's
optimal strategy is to use a completely entangled quantum notepad, on
which to encode his prior information. 
\end{abstract}

\maketitle

\section{Introduction}

The well-known classical Monty Hall problem, also known under
various other names \cite{ctk}, is set in the context of a
television game show. It can be seen as a two person game, in
which a player P tries to win a prize, but a show master (or Quiz
master) Q tries to make it difficult for her \cite{feminine}. Like
other games of this sort \cite{Wilkens1,Wilkens2} it can be
``quantized'', i.e., its key elements can be formulated in a
quantum mechanical context, allowing new strategies and new
solutions. However, such quantizations are rarely unique, and
depend critically on what is seen as a ``key element'', and also
on how actions which might change the system are formalized, 
corresponding to how, in the classical version, 
information is gained by ``looking at something''. 
The Monty Hall problem is no exception, and there are 
already quantizations \cite{Hefei,FA}. The version we
present in this paper was drafted independently, and indeed we
come to a quite different conclusion. We discuss the relation
between these two approaches and ours in more detail in
Sec.~\ref{s:variants} below.

Our paper is organized as follows. In Sec.~\ref{s:classical} we
review the classical game, whose various steps are translated to
the quantum context in Sec.~\ref{s:quantum}. In
Sec.~\ref{s:variants} we discuss some alternative ideas to
quantize the game. As in the classical version, computer
simulations help to understand the structure of the game and its
basic strategic options. We provide a Java simulation
\cite{applet} of our version from the point of view of player P.
There is a basic strategy for P in our version, which we call the
{\it classical strategy} (Sec.~\ref{s:classtrat}), which
guarantees a gain for her at least as good as in the classical
game. The further analysis depends on how Q records the
information about the initial preparation, which he needs later on
in the game: if he uses a classical notepad, he has to choose his
strategy very carefully, or P can improve her odds beyond the
classical expectation (Sec.~\ref{s:beat}). These possibilities for
P are illustrated especially nicely by the simulation
\cite{applet}. The case of a quantum notepad is discussed in
Sec.~\ref{s:entangle}, and gives Q a simpler way to restrict P's
expectations to the classical value.

\section{The Classical Monty Hall Problem}\label{s:classical}

The classical Monty Hall problem is set in the context of a
television game show. In the last round of the show, the
candidates were given a chance to collect their prize (or lose it)
in the following game:
 \begin{enumerate}
 \item Before the show the prize is hidden behind one of three closed doors.
The show master knows where the prize is but, of course, the
candidate does not.
 \item The candidate is asked to choose one of the three doors, which is,
however, not opened at this stage.
 \item The show master opens another door, and shows that there is
no prize behind it. (He can do this, because he knows where the
prize is).
 \item The candidate can now open one of the remaining doors to
either collect her prize or lose.
 \end{enumerate}
 Of course, the question is: should the candidate stick to her
original choice or ``change her mind'' and pick the other
remaining door? As a quick test usually shows, most people will
stick to their first choice. After all, before the show master
opened a door the two doors were equivalent, and they were not
touched (nor was the prize moved). So they should still be
equivalent. This argument seems so obvious that trained
mathematicians and physicists fall for it almost as easily as
anybody else \cite{embarass}.

However, the correct solution by which the candidates can, in
fact, double their chance of winning, is to always choose the
other door. The quickest way to convince people of this is to
compare the game with another one, in which the show master offers
the choice of either staying with your choice or {\it opening
both other doors}. Anybody would prefer that, especially if the
show master courteously offers to open one of the doors for you.
But this is precisely what happens in the original game when you
always change to the other door.

\section{The Quantum Monty Hall Problem}\label{s:quantum}

We will ``quantize'' only the key parts of the problem. That is,
the prize and the players, as well as their publicly announced
choices, will remain classical. The quantum version can even be
played in a game show on classical TV. 

The main quantum variable will be the position of the prize. It lies in a
$3$-dimensional complex Hilbert space $\H$, called the {\it game
space}. We assume that an orthonormal basis is fixed for this
space so that vectors can be identified by their components, but
apart from this the basis has no significance for the game.
A second important variable in the game is what we will call 
the show master's notepad. This might be classical information
describing how the game space was prepared, or it might be a quantum
system, entangled with the prize. In the latter case, the show master
is able to do a quantum measurement on his notepad, 
providing him with classical information about the prize, 
without moving the prize, in the sense
that the player's information about the prize is not changed by the
mere fact that the show master ``consults his notepad''.  
A measurement on an auxiliary quantum system, even
if entangled with a system of interest, does not alter the reduced state
of the system of interest. After the show master has consulted his
notepad, we are in the same situation as if the notepad had been a
classical system all along. As in the classical game, 
the situation for the player might change 
when the show master, by opening a door, reveals to some extent
what he saw in his notepad. 
Opening a door corresponds to a measurement along a one
dimensional projection on $\H$. 

The game proceeds in the
following stages, closely analogous to the classical game:
 \begin{enumerate}
 \item Before the show the game space system is prepared quantum
mechanically. Some information about this preparation is given to
the show master Q. This can be in the form of another system,
called the notepad, which is in a state correlated with the game
space.
 \item The candidate chooses some one dimensional projection $p$ on $\H$.
 \item The show master opens a door, i.e., he chooses a one
dimensional projection $q$, and makes a L{\"u}ders/von Neumann
measurement with projections $q$ and $(\idty-q)$.  In order to do
this, he is allowed first to consult his notebook. If it is a quantum system,
this means that he carries out a measurement on the notebook.
The joint state of prize and notebook then change, but the traced
out or reduced state of the prize does not change, as far as the player
is concerned. Two rules constrain the show master's choice of $q$: 
he must choose ``another door'' in the sense
that $q\perp p$; and he must be {\it certain not to reveal the
prize}. The purpose of his notepad is to enable him to do this.  After these
steps, the game space is effectively collapsed to the two-dimensional
space $(\idty-q)\H$. 
 \item The player P can now choose a one dimensional projection
$p'$ on $(\idty-q)\H$, and the corresponding measurement on the
game space is carried out. If it gives ``yes'' she collects the
prize.
 \end{enumerate}
As in the classical case, the question is: how should the player
choose the projection $p'$ in order to maximize her chance of
winning? Perhaps it is best to try out a few options in a
simulation, for which a Java applet is available \cite{applet}.
For the input to the applet, as well as for some of the discussion
below it is easier to use unit vectors rather than one-dimensional
projections. As standard notation we will use for $p=\kettbra\Phi$
for the door chosen by the player, $q=\kettbra\chi$ for the door
opened by $Q$, and $r=\kettbra\Psi$ for the initial position of
the prize, if that is defined.

 From the classical case it seems likely that choosing $p'=p$ is
a bad idea. So let us say that the {\it classical
strategy} in this game consists of always switching to the
orthogonal complement of the previous choice, i.e., to take
$p'=\idty-q-p$. Note that this is always a projection because, by
rule 3, $p$ and $q$ are orthogonal one dimensional projections. We
will analyze this strategy in Sec.~\ref{s:classtrat}, which turns
out to be possible without any specification of how the show master 
can guarantee not to stumble on the prize in step~3.

There are two main ways the show master can satisfy the rules.
The first is that he chooses randomly the components of a vector in 
$\H$, and prepares the game space in the corresponding pure state. 
He can then just take a note of his choice on a classical pad, so that in
stage~3 he can compute a vector orthogonal to both the direction
of the preparation and the direction chosen by the player. Q's
strategies in this case are discussed in Sec.~\ref{s:classtrat}.
The second and more interesting way is to use a quantum
notepad, i.e., another system with three dimensional Hilbert space
$\N$, and to prepare a ``maximally entangled state'' on
$\H\otimes\N$. Then until stage~3 the position of the prize is
completely undetermined in the strong sense only possible in
quantum mechanics, but the show master can find a safe door to 
open on $\H$ by
making a suitable measurement on $\N$. Q's strategies in this case
are discussed in Sec.~\ref{s:entangle}.

\section{The classical strategy}\label{s:classtrat}

To explain why the classical strategy works almost as in the
classical version of the problem, we look more closely at the
situation, when Q has just reached the decision which door $q$ he
wants to open. His method of arriving at
that decision is irrelevant. It must take into account the player's 
choice $p$, as well as  Q's  information about the preparation, 
and may involve some further randomness or even a quantum measurement
on an auxiliary quantum system called the notepad.  
The choice of $q$ is announced publicly
by ``opening'' that door, and needs to be known as classical
information to P for her to choose a door $p'$ orthogonal to it.
So we shall consider $q$ as a classical random variable. What we
need to study are the correlations of this variable with outcomes of
later measurements made on the game system. That is to say, 
we have a
{\it hybrid system} described by a combination of classical and
quantum variables.

There is a standard form for states on such systems, which we will
need to use here. Let us consider first the classical part. Since $p$
was arbitrary and since the procedure used by the show master might
involve classical randomization or quantum randomness (through
measurement of a quantum notepad), $q$ has a probability distribution,
denoted by $w$, which typically will depend on $p$.  Thus the
expectation of any real valued function $f$ of $q$ is to be computed
as $\int\!w(dq)\;f(q)$. Next we are interested in the joint response of a
classical detector (described by a characteristic function $f$) and a
quantum detector, e.g., given by a one dimensional projection $p'$. By
the theorem of Ozawa\cite{ozawa} the additional information needed to
compute such expectations is encoded by \emph{conditional density
  operators} $\rho_q$. Then 
\begin{equation}
  \int\!w(dq)\;\tr(\rhoq p')f(q)
\end{equation}
is the joint response probability for $f$ and $p'$. This too will
typically depend on $p$, but we suppress that dependence also from the
notation. Loosely speaking, $\rho_q$ is the density matrix which $P$ has
to use for the game space after Q has announced his intention of
opening door $q$. This is usually not the same conditional density
operator as the one used by Q: Since Q has more classical information
about the system, he may condition on that, leading to finer
predictions. In contrast, $\rhoq$ is conditioned only on the publicly
available information. 

From $w$ and $\rhoq$ we can compute the marginal density operator for
the quantum subsystem, describing measurements without consideration
of the classical variable $q$. This is the mean density
operator
\begin{equation}\label{rhom}
  \rhom=\int\!w(dq)\;\rhoq\;.
\end{equation}
It will not depend on $p$ and it will be the same as the reduced
density operator for the game space before the show master consults his
notepad (he is not allowed to touch the prize),
and even before the player chooses $p$ (which cannot affect the prize).

From the rules alone we know two things about the conditional density
operators: firstly, that $\tr(\rhoq\;q)=0$: the show master must
not hit the prize. Secondly, $q$ and $p$ must commute, so it does
not matter which of the two we measure first. Thus a measurement of
$p$ responds with probability
$\int\!w(dq)\;{\tr(\rhoq\;p)}=\tr(\rhom\;p)$. Combining these two we get
the overall probability $w_c$ for winning with the classical strategy
as 
\begin{equation}\label{e:class}
  w_c=\int\!w(dq)\;{\tr\bigl(\rhoq(\idty-p-q))}
     =1-\tr(\rhom\;p)\;.
\end{equation}

If we assume that $\rhom$ is known to P, from watching the show
sufficiently often, the best strategy for P is to choose initially
the $p$ with the smallest expectation with respect to $\rhom$,
just as in the classical game with uneven prize distribution it
is best to choose initially the door least likely to contain the
prize. If $Q$ on the other hand wants to minimize P's gain
\cite{ratings}, he will choose $\rhom$ to be uniform, which in the
quantum case means $\rhom=\frac13\idty$, and hence $w_c=2/3$.

\section{Strategies against classical notepads}\label{s:beat}

In this section we consider the case that the show master records
the prepared direction of the prize on a classical notepad. We
will denote the one dimensional projection of this preparation by
$r$. Then when he has to open a door $q$, he needs to choose
$q\perp r$ and $q\perp p$. This is always possible in a three
dimensional space. But unless  $p=r$, he has no choice: $q$ is
uniquely determined. This is the same as in the classical case,
only that the condition ``$p=r$'', i.e., that the player chooses
{\it exactly} the prize vector typically has probability zero.
Hence Q's strategic options are not in the choice of $q$, but
rather in the way he randomizes the prize positions $r$, i.e., in
the choice of a probability measure $v$ on the set of pure states.
In order to safeguard against the classical strategy he will make
certain that the mean density operator $\rhom=\int\!v(dr)\; r$ is
unpolarized ($=\frac13\idty$). It seems that this is about all he
has to do, and that the best the player can do is to use the
classical strategy, and win $2/3$ of the time. However, this turns
out to be completely wrong.

\subsection{Preparing along the axes}

Suppose the show master decides that since the player can win as
in the classical case, he might as well play classical as well,
and save the cost for an expensive random generator. Thus he fixes
a basis and chooses each one of the basis vectors with probability
$1/3$. Then $\rhom=\frac13\idty$, and there seems to be no
giveaway. In fact, the two can now play the classical version,
with P choosing likewise a projection along a basis vector.

But suppose she does not, and chooses instead the projection along
the vector $\Phi=(1,1,1)/\sqrt3$. Then if prize happens to be
prepared in the direction $\Psi=(1,0,0)$, the show master has no
choice but to choose for $q$ the unique projection orthogonal to
these two, which is  along $\chi=(0,1,-1)$. So when Q announces
his choice, P only has to look which component of the vector is
zero, to {\it find the prize with certainty!}

This might seem to be an artifact of the rather minimalistic
choice of probability distribution. But suppose that $Q$ has
settled for any {\it arbitrary finite collection of vectors}
$\Psi_\alpha$ and their probabilities. Then P can choose a vector
$\Phi$ which lies in none of the two dimensional subspaces spanned
by two of the $\Psi_\alpha$. This is possible, even with a random
choice of $\Phi$, because the union of these two dimensional
subspaces has measure zero. Then, when Q announces the projection
$q$, P will be able to reconstruct the prize vector with
certainty: at most one of the $\Psi_\alpha$ can be orthogonal to
$q$. Because if there were two, they would span a two dimensional
subspace, and together with $\Phi$ they would span a three
dimensional subspace orthogonal to $q$, which is a contradiction.

Of course, any choice of vectors announced with floating point
precision is a choice from a finite set. Hence the last argument
would seem to allow P to win with certainty in every realistic
situation. However, this only works if she is permitted to ask for
$q$ at any desired precision. So by the same token (fixed length
of floating point mantissa) this advantage is again destroyed.

This shows, however, where the miracle strategies come from: by
announcing $q$, the show master has not just given the player
$\log_23$ bits of information, but an infinite amount, coded in
the digits of the components of $q$ (or the vector $\chi$).

\subsection{Preparing real vectors}

The discreteness of the probability distribution is not the key
point in the previous example. In fact there is another way to
economize on random generators, which proves to be just as
disastrous for Q. The vectors in $\H$ are specified by three
complex numbers. So what about choosing them real for simplicity?
An overall phase does not matter anyhow, so this restriction does
not seem to be very dramatic.

Here the winning strategy for P is to take $\Phi=(1,i,0)/\sqrt2$,
or another vector whose real and imaginary parts are linearly
independent. Then the vector $\chi\perp\Phi$ announced by Q will
have a similar property, and also must be orthogonal to the real
prize vector. But then we can simply compute the prize vector as
the outer product of real and imaginary part of $\chi$.

For the vector $\Phi$ specified above we find that if the prize is
at $\Psi=(\Psi_1,\Psi_2,\Psi_3)$, with $\Psi_k\in\Rl$, the unique
vector $\chi$ orthogonal to $\Phi$ and $\Psi$ is connected to
$\Psi$ via the transformations
\begin{eqnarray}
  \chi&\propto&(\Psi_3,\;-i\Psi_3,\;-\Psi_1+i\Psi_2) \label{Psi2chi}\\
  \Psi&\propto&(-\Ree\chi_3,\;\Imm\chi_3,\;\chi_1)\;, \label{chi2Psi}
\end{eqnarray}
where ``$\propto$'' means ``equal up to a factor'', and it is
understood that an overall phase for $\chi$ is chosen to make
$\chi_1$ real. This is also the convention used in the simulation
\cite{applet}, so Eq. (\ref{chi2Psi}) can be tried out as a
universal cheat against show masters using only real vectors.

\subsection{Uniform distribution}
 The previous two examples have one thing in common: the probability
distribution of vectors employed by the show master is
concentrated on a rather small set of pure states on $\H$.
Clearly, if the distribution is more spread out, it is no longer
possible for P to get the prize every time. Hence it is a good
idea for Q to choose a distribution which is as uniform as
possible. There is a natural definition of ``uniform''
distribution in this context, namely the unique probability
distribution on the unit vectors, which is invariant under
arbitrary unitary transformations \cite{uniform}. Is this a good
strategy for Q?

Let us consider the conditional density operator $\rhoq$, which
depends on the two orthogonal projections $p,q$. It implicitly
contains an average over all prize vectors leading to the same
$q$, given $p$. Therefore, $\rhoq$ must be invariant under all
unitary rotations of $\H$ fixing these two vectors, which means
that it must be diagonal in the same basis as $p,q,(\idty-p-q)$.
Moreover, the eigenvalues cannot depend on $p$ and $q$, since
every pair of orthogonal one dimensional projections can be
transformed into any other by a unitary rotation. Since we know
the average eigenvalue in the $p$-direction to be $1/3$, we find
\begin{equation}\label{rhoquniform}
  \rhoq=\frac13p+\frac23(\idty-p-q)\;.
\end{equation}
Hence the classical strategy for P is clearly optimal. In other
words, the pair of strategies: ``uniform distribution for Q and
classical strategy for P'' is an equilibrium point of the game. We do
not know yet, whether this equilibrium is unique, in other words: If Q
does not play precisely by the uniform distribution: can P always
improve on the classical strategy? We suppose that the answer to
this question is yes; to find a proof of this conjecture has turned
out, however, to be a hard problem which is still open.

\section{Strategies for Quantum notepads}\label{s:entangle}
\subsection{Quantum notepad and entanglement}
As briefly described in Sec.~\ref{s:quantum}, a notepad is quite
generally another physical system, which by its initial
preparation is brought into a state correlated with the game
system, so that measurements on the notepad provide information
about the game system as well. The classical notepad we were
considering in the previous section can be formalized in this way,
too, although it may appear a bit artificial to introduce a
physical system on which some random variable is ``written''. For
the quantum notepads considered in this section, however, this
view is mandatory.

The simplest form of a quantum notepad meeting Q's requirements is
another system with three dimensional Hilbert space $\N$. Denoting
some bases in $\H$ and $\N$ by ket vectors $\ket1,\ket2,\ket3$,
and abbreviating $\ket
k\otimes\ket\ell\equiv\ket{k\ell}\in\H\otimes\N$, we introduce the
``maximally entangled vector''
\begin{equation}\label{maxent}
  \Omega=\frac1{\sqrt3}\sum_{k=1}^3\ket{kk}\;.
\end{equation}
It clearly depends on the bases we have chosen, but not as much as
one might expect. Its crucial property is the equation
\begin{equation}\label{maxtrans}
  (X\otimes\idty)\Omega=(\idty\otimes X^T)\Omega\;,
\end{equation}
where the transpose $X^T$ is defined by $\bra
kX^T\ket\ell=\bra\ell X\ket k$. Note that in this way an operator
$X$ on $\H$ becomes an operator $X^T$ on $\N$, and this becomes
possible by using the same labels for the basis vectors in both
spaces. If we now change the basis in $\H$ by a unitary operator
$U$, say, we find that there is a corresponding basis change on
$\N$, leaving the vector $\Omega$ invariant, namely
$(U^{-1}\otimes U^T)\Omega=\Omega$. Hence $\Omega$ does not depend
on the individual bases, but represents a particular way of
identifying $\H$ with $\N$ \cite{ccHilbert}, and all bases
accordingly. This is underlined by the observation that the
restriction of this state to either factor is the completely
unpolarized state, i.e.,
\begin{equation}\label{restrOmega}
  \bra\Omega A\otimes\idty\ket\Omega
  =\bra\Omega\idty \otimes A\ket\Omega
  =\frac13\tr(A)\;,
\end{equation}
i.e., the unique state invariant under all unitary rotations.

Let us consider now a measurement on the notepad. It is described,
like any measurement, by a positive operator valued measure. Let
us take it discrete-valued for the moment to simplify the
explanation. Thus the measurement is given by a collection of
positive operators $F_x$ on $\N$, such that $\sum_xF_x=\idty$.
Here the labels ``$x$'' are the (classical) outcomes of the
measurement, and $\tr(\rho\;F_x)$ is interpreted as the
probability of getting this result, when measuring on systems
prepared according to $\rho$. How could Q now infer from this
result a safe door $q$ for him to open in the game? This would
mean that $F_x$ measured on $\N$, and $q$ measured on $\H$ never
give a simultaneous positive response, when measured in the state
$\Omega$, i.e.,
\begin{equation}\label{qperpF}
  0=\bra\Omega q\otimes F_x\ket\Omega
   =\bra\Omega \idty\otimes F_xq^T\ket\Omega
   =\frac13\tr(q^T\;F_x)\;.
\end{equation}
Since $F_x$ and $q^T$ are both positive, this equivalent to
$F_xq^T=0$. Of course, Q's choice must also satisfy the constraint
$q\perp p$. There are different ways of arranging this, which we
discuss in the following two subsections.

\subsection{Equivalence if observable is chosen beforehand}
Suppose Q chooses the measurement beforehand, and let us suppose
it is discrete, as before. Then for every outcome $x$ and every
$p$ he must be able to find a one dimensional projection
satisfying both constraints $F_x^Tq=0$ and $qp=0$. Clearly, this
requires that $F_x$ has at least a two dimensional null space,
i.e., $F_x=\kettbra{\phi_x}$, with a possibly non-normalized
vector $\phi_x\in\N$. It will be convenient to take the vectors
$\phi_x$ to be normalized, and to define $F_x=v_x\kettbra{\phi_x}$
with factors $v_x$ summing to $3$, the dimension of the Hilbert
space (take the trace of the normalization condition for $F$). We
can further simplify this structure, by identifying outcomes $x$
with the same $\phi_x$, since for these the same projection $q$
has to be chosen anyhow. We can therefore drop the index ``$x$'',
and consider the measure to be defined directly on the set of one
dimensional projections. But this is precisely the structure we
had used to describe a classical notepad. This is not an
accidental analogy: apart from taking transposes as in
equation~(\ref{maxtrans}) this measure has precisely the same
strategic meaning as the measure of a classical notepad.

This is not surprising: if the observable is chosen beforehand, it
does not matter whether the show master actually performs the
measurement before or after the player's choice. But if he does it
before P's choice, we can just as well consider this measurement
with its classical output as part of the preparation of a
classical notepad, in which the result is recorded.

\subsection{Simplified strategy for Q}

Obviously the full potential of entanglement is used only, when Q
chooses his observable after P's choice. Since the position of the
prize is ``objectively undetermined'' until then, it might seem
that there are now ways to beat the 2/3 limit. However, the
arguments for the classical strategy hold in this case as well. So
the best Q can hope for are some simplified strategies. For
example, he can now get away with something like measuring along
axes only, even though for classical notepads using ``axes only''
was a certain loss for Q.

We can state this in a stronger way, by introducing tougher rules
for Q: In this variant P not only picks the direction $p$, but
also two more projections $p'$ and $p''$ such that
$p+p'+p''=\idty$. Then Q is not only required to open a door
$q\perp p$, but we require that either $q=p'$ or $q=p''$. It is
obvious how Q can play this game with an entangled notepad: he
just uses the transposes of $p,p',p''$ as his observable. Then
everything is as in the classical version, and the equilibrium is
again at $2/3$.

\section{Alternative versions and quantizations of the
game}\label{s:variants}

\subsection{Variants arising already in the classical case}
\label{sec:vari-aris-alre}
Some variants of the problem can also be considered in the
classical case, and they tend to trivialize the problem, so that
P's final choice becomes equivalent to ``Q has prepared a coin,
and P guesses heads or tails''. Here are some possibilities,
formulated in a way applying both to  the classical and the
quantum version.
 \begin{itemize}
 \item{\it Q is allowed to touch the prize after P made her first
choice}. Clearly, in this case Q can reshuffle the system, and
equalize the odds between the remaining doors. So no matter what P
chooses, there will be a 50\% chance for getting the prize.

 \item{\it Q is allowed to open the door first chosen by P}. Then
there is no way P's first choice enters the rules, and we may
analyze the game with stage 2 omitted, which is entirely trivial.

 \item{\it Q may open the door with the prize, in which case the game
starts again}. Since Q knows where the prize is, this is the same
as allowing him to abort the round, whenever he does not like what
has happened so far, e.g., if he does not like the relative
position of prize and P's choice. In the classical version he
could thus cancel 50\% of the cases, where P's choice is not the
prize, thus equalizing the chances for P's two pure strategies.
Similar possibilities apply in the quantum case.

\end{itemize}

\subsection{Variants in which classical and quantum behave differently}

\begin{itemize}
 \item{\it Q may open the door with the prize, in which case P
gets the prize}. In the classical version, revealing the prize is
then the worst possible pure strategy, so mixing in a bit of it
would seem to make things always worse for Q. Then although
increasing Q's options in principle can only improve things for
$Q$, one would advise him not to use the additional options. This
is assuming, though, that in the remaining cases Q sticks to his
old strategy. However, even classically, the relaxed rule gives
him some new options: He can simply ignore the notepad, and open
any door other than $p$. Then the game becomes effectively ``P and
Q open a door each, and P gets all prizes''. Assuming uniform
initial distribution of prizes this gives the same $2/3$ winning
chance as in the original game.

The corresponding quantum strategy works in the same way.
Assuming, for simplicity,  a uniform mean density operator
$\rhom=\frac13\idty$, Q's strategy of ignoring his prior
information will give the classical $2/3$ winning chance for P.
But this is a considerable improvement for Q in cases where a
non-uniform probability distribution of pure states previously
gave Q a 100\% chance of winning. So in the quantum case, doing
two seemingly stupid things together amounts to a good strategy
for Q: firstly, sometimes revealing the prize for P, and secondly
ignoring all prior information.

Note that this strategy is optimal for Q, because the classical
strategy still guarantees the $2/3$ winning chance for P. This can be
seen with the same arguments as in Section \ref{s:classtrat}. The only
difference is that $\tr(\rho_q q)$ can be nonzero, since Q may open the
door with the prize. However in this case P wins and we get instead of
Equation (\ref{e:class})
\begin{eqnarray}
  w_c &=& \int\!w(dq)\;{\tr\bigl(\rhoq(\idty-p-q))} + \tr(\rho_q q) \\
  &=& 1 - \tr(\rhom p) = \frac{2}{3}
\end{eqnarray}

\item{\it As Q opens the door he is allowed to make a complete von Neumann
measurement}. Classically, it would make no difference if the
doors were completely transparent to the show master. He would not
even need a pad then, because he could always look where the prize
is. But ``looking'' is never innocent in quantum mechanics, and in
this case it is tantamount to moving the prize around. So let us
make it difficult for Q, by insisting that the initial preparation
is along a fixed vector, known also to P, and that Q not only has
to announce the direction $q$ of the door he opens, but also the
projections $q'\perp q$ and $q''=\idty-q-q'$ entering in the
complete von Neumann measurement, which takes an arbitrary density
operator $\rho$ to
\begin{equation}\label{completevN}
  \rho\mapsto q\rho q+q'\rho q'+q''\rho q''.
\end{equation}
 Moreover, we require as before, that $q$ is orthogonal both to
$p$ and to the prize. The only thing remaining secret is which of
the projections $q'$ and $q''$ has detected the presence of the
prize (This simply would allow P to open that door and collect).
Q's simple strategy is now to choose $q$ as before. The position
of $p$ is irrelevant for his choice of $p'$ and $p''$: he will
just take these directions at $45^\circ$ to the prize vector. This
will result in the unpolarized density operator $(q'+q'')/2$, and
no matter what P chooses, her chances of hitting the prize will be
$1/2$. She will probably feel cheated, and she is, because even
though she knows exactly where the prize was initially, the
strategy ``choose the prize, and stick to this choice'' no longer
works.

\end{itemize}

\subsection{Two published versions}

\begin{itemize}
\item 
The quantization proposed in \cite{Hefei} is probably \cite{Hefei2}
closely related to the second variant of the last Subsection: After Q
has opened one door he is allowed to perform an arbitrary von Neumann
measurement on the  remaining two-dimensional subspace -- i.e. he
``looks'' where the prize is. In the classical case this is an allowed
(but completely superfluous) step. In the quantum case, however, the
prize is shuffled around and we end up with a variant from Subsection 
\ref{sec:vari-aris-alre}: $Q$ is allowed to move the prize. In other
words, the final result of the game is completely  independent of the
steps prior to this measurement and the whole game is reduced to coin
tossing -- which is not very interesting.
\item 
A completely different quantization of the game is given in
\cite{FA}. In contrast to our approach, the moves available to Q and
P are here not preparations and mesurements but \emph{operations} on a
tripartite system which is initially in the pure state $\psi \in \H_Q
\otimes \H_P \otimes \H_O$ (and different choices for $\psi$ lead to
different variants of the game). The Hilbert spaces $\H_Q,
\H_P$ and $\H_O$ describe the doors where $Q$ hides the
prize, which P chooses in the second step and which Q opens
afterwards and 
the gameplay is described by the unitary operator 
\begin{equation}
  U = (\cos \gamma U_S + \sin \gamma U_N) U_O (U_Q \otimes U_P \otimes \idty).
\end{equation}
$U_Q$ and $U_P$ are arbitrary unitaries, describing Q's and
$P$'s initial choice, $U_O$ is the (fixed) opening box operator and
$U_S$ respectively $U_N$ are P's ``switching'' and ``not-switching''
operators. The payoff is finally given as the expectation value of
an appropriate observable $\$$ (for a precise definition of $U_O$, $U_S$,
$U_N$ and $\$$ see \cite{FA}). The basic idea behind this scheme is
quite different from ours and a comparison of results is therefore
impossible. Nevertheless, this is a nice example which shows that
quantizing a classical game is very non-unique.

\end{itemize}

\section*{Acknowledgments}
The idea for this paper came up in a bar during a very stimulating
workshop ``Qrandom'', conducted at the research center EURANDOM at
Eindhoven, Feb. 12-16, 2001.  Financial support by the center, is
gratefully acknowledged. We whish to thank also M. Guta for useful
discussion about the uniqueness problem discussed at the end of
Section \ref{s:beat}.


\begin{references}
\bibitem{feminine}In this text, the show master is male, like
Monty Hall, the host of the show, where the game first appeared.
The player is female, like Marilyn vos Savant, who was the first
to fight the public debate for the recognition of the correct
solution, and had to take some sexist abuse for that.
\bibitem{embarass}The only difference seems to be that
professionals tend to pontificate about the obvious truth of the
false solution. Embarassingly many have gone on record.
\bibitem{ratings}Of course, real game show hosts don't care,
they try to maximize ratings. Incidentally, this means that the
quantum version will hardly ever be played on classical TV. For
how can a black box containing some esoteric game space system
compete with the dramatic potential of closed doors?
\bibitem{uniform}Numerically the way to generate this distribution
is to take the real and imaginary parts of all components of a
vector as independent identically normally distributed random
variables, and to normalize the resulting vector.
\bibitem{ccHilbert}More precisely, $\Omega$ is equivalent to an
identification of $\H$ with the complex conjugate Hilbert space of
$\N$. This structure routinely arises in the so-called
GNS-representation of operator algebras.
\bibitem{Hefei2}This is only one possible reading of \cite{Hefei},
  since the authors do not specify the rules of their game
  with sufficient clarity.
\end{references}
\end{document}